\documentclass[11pt]{article}
\usepackage[margin=1in]{geometry}
\usepackage{graphicx}

\def\Title#1{\begin{center} {\Large {\bf #1} } \end{center}}
\def\Author#1{\begin{center} {\normalsize {\sc #1} } \end{center}}
\def\Institution#1{\begin{center} {\normalsize {\it #1} } \end{center}}
\def\Abstract#1{\noindent {\normalsize {\bf Abstract:} {\normalfont #1}}}
\def\Conference{\vspace{4mm}\begin{raggedright} {\normalsize {\it Poster presented at the 2019 Meeting of the Division of Particles and Fields of the American Physical Society (DPF2019), July 29--August 2, 2019, Northeastern University, Boston, C1907293.} } \end{raggedright}\vspace{4mm}}

\begin{document}

%
%

\Title{Fabrication of a Cosmic Ray Veto System for the Mu2e Experiment}

\Author{H. Woodward}

\Institution{Department of Physics\\ University of Virginia, 22904 VA, USA}

\Conference{}

\Abstract{The Mu2e experiment at Fermilab will search for the charged-lepton flavor-violating process of a neutrino less muon-to-electron
  decay in the presence of a nucleus. The experiment expects a
  single-event sensitivity of $2.9\times10^{-17}$, which is four
  orders of magnitude below the current strongest limits on this process. This requires all backgrounds to sum to fewer than one event over the lifetime of the experiment. One major background is due to cosmic-ray muons producing electrons that fake a signal inside of the Mu2e apparatus. The Mu2e Cosmic Ray Veto (CRV) has been designed to veto these cosmic-ray backgrounds with an efficiency of 99.99\%, while causing a low dead time and while operating in a high-intensity environment. The design and fabrication of the CRV is discussed.}

\section{The Mu2e Experiment}
The Mu2e experiment at Fermilab will search for a neutrino less muon-to-electron conversion in the presence of an aluminum nucleus. The observance of such a conversion would be clear evidence of physics beyond the standard model. The failure to observe such a conversion would still be a useful constraint to possible standard model extensions.  

In the total run-time of Mu2e, it will be exposed to approximately
$10^{18}$ muons generated from a proton beam, resulting in a
single-event sensitivity of $2.9\times10^{-17}$ \cite{CLFV}. This is
10,000 times better than the current limit \cite{SINDRUMII}. However,
a major background source are the over 15,000 cosmic-ray muons that
are expected to pass through the detector every second, producing a
signature-like event about once per day. In order to suppress this
cosmic-ray background to below 0.25 events over the Mu2e runtime
\cite{TDR}, the Cosmic Ray Veto has been designed to have an
efficiency of at least 99.99\%. It is expected to cause an an experimental dead time
of less than 10\% for Mu2e due to the neutrons and gammas coming from the muon beamline. 

\begin{figure}[htb]
\centering
   \includegraphics[scale=0.4]{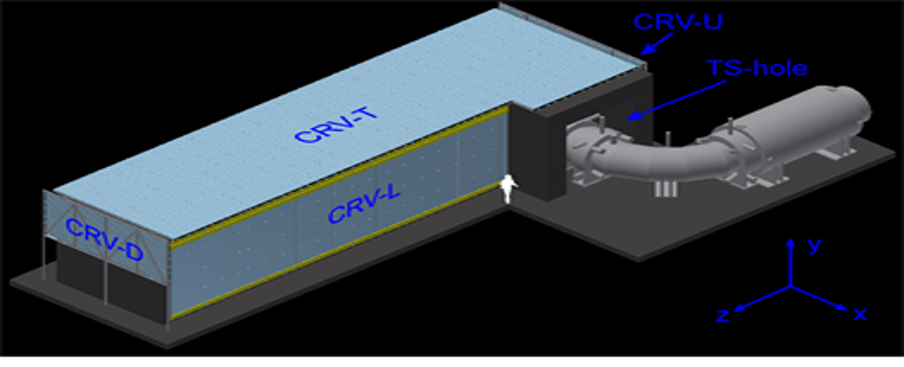}
    \caption{The CRV encloses the detector solenoid, which surrounds the stopping target, the tracker, and the calorimeter of the Mu2e experiment.}
    \label{fig:my_label0}
\end{figure}

\section{The Cosmic Ray Veto}
The Mu2e Cosmic Ray Veto (CRV) encloses the detector solenoid and part of the transport solenoid in order
to identify background cosmic rays (as shown Fig.~\ref{fig:my_label0}). The design of the CRV is driven by the need for excellent muon veto efficiency, large area, small gaps, low cost, access to electronics, and operation in a strong magnetic field.

\subsection{Di-counter Construction}
\begin{figure}[htb]
 \centering
    \includegraphics[scale=0.5]{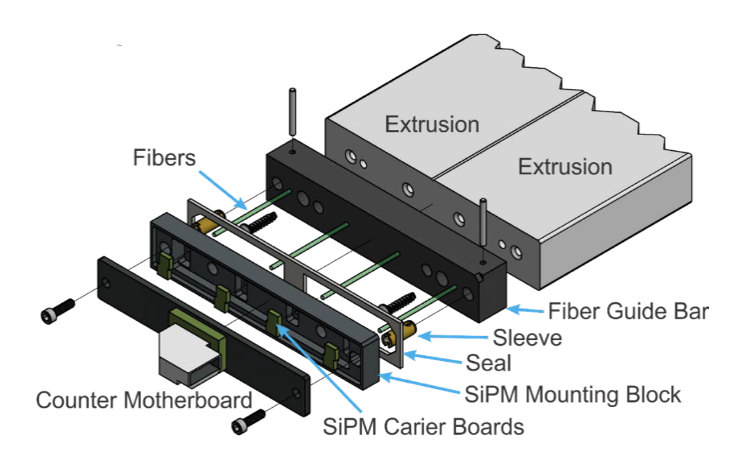}
    \caption{Di-counter components.}
    \label{fig:my_label1}
\end{figure}

The fundamental component of the CRV is a extruded polystyrene
scintillator counter coated with titanium dioxide, with lengths ranging from 1.0 to 6.9 meters. Two of these counters are joined together
with epoxy to form a di-counter. Wavelength-shifting
fibers~\cite{FiberPaper} are then fed through each of four channels in the two extrustions. Fiber Guide Bars (FGBs) are attached to the
scintillator face in order to precisely align the fibers to the readout electronics. Silicon photomultipliers (SiPMs)
detect the light from fibers. During fabrication, the faces of the
FGBs are flycut to polish the surface for optimal light transmission
to the readout.  The components of a di-counter used to couple to the
readout electronics are shown in Fig.~\ref{fig:my_label1}.

\subsection{Di-counter Quality Assurance}
Each constructed di-counter undergoes a series of tests to ensure that its performance is adequate for a high-efficiency CRV. 

First, the ends of each di-counter are photographed and examined in order to detect any scratches, cracks, or other damage to the fiber surface. Sometimes a di-counter with a damaged fiber end can be saved by re-flycutting it. A roughness tester is used to monitor for signs of wear on the cutting bit by analyzing the roughness of the flycut FGB surface. Next, LEDs are flashed through each fiber end. This transmission test reads the light yield in order to detect any internal damage to the fiber.

\begin{figure}[htb]
    \centering
    \includegraphics[scale=0.35]{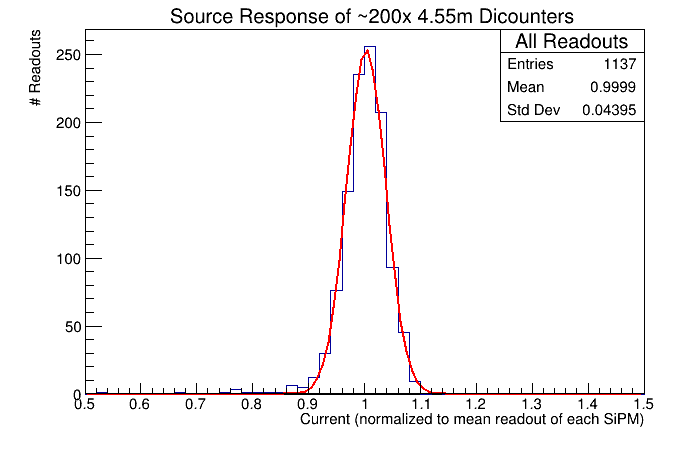}
    \caption{Normalized source response of $\approx200$ dicounters. The relative values of each SiPM readout to the mean value of the corresponding SiPM over all measurements (each mean normalized to be one) are calculated and shown plotted together.}
    \label{fig:my_label2}
\end{figure}

Finally, each di-counter is placed inside of a dark box for source
testing. A Cs-137 source is positioned over the di-counter one meter
from each end and the induced current is measured. If any aspect of a
di-counter is damaged, it becomes apparent with lower current readouts
at this stage. A di-counter with a response under 80\% of the nominal
is rejected.  A typical distribution of source
measurements for about 200 di-counters is shown in Fig.~\ref{fig:my_label2}.

\begin{figure}[htb]
  \centering
    \includegraphics[scale=0.30]{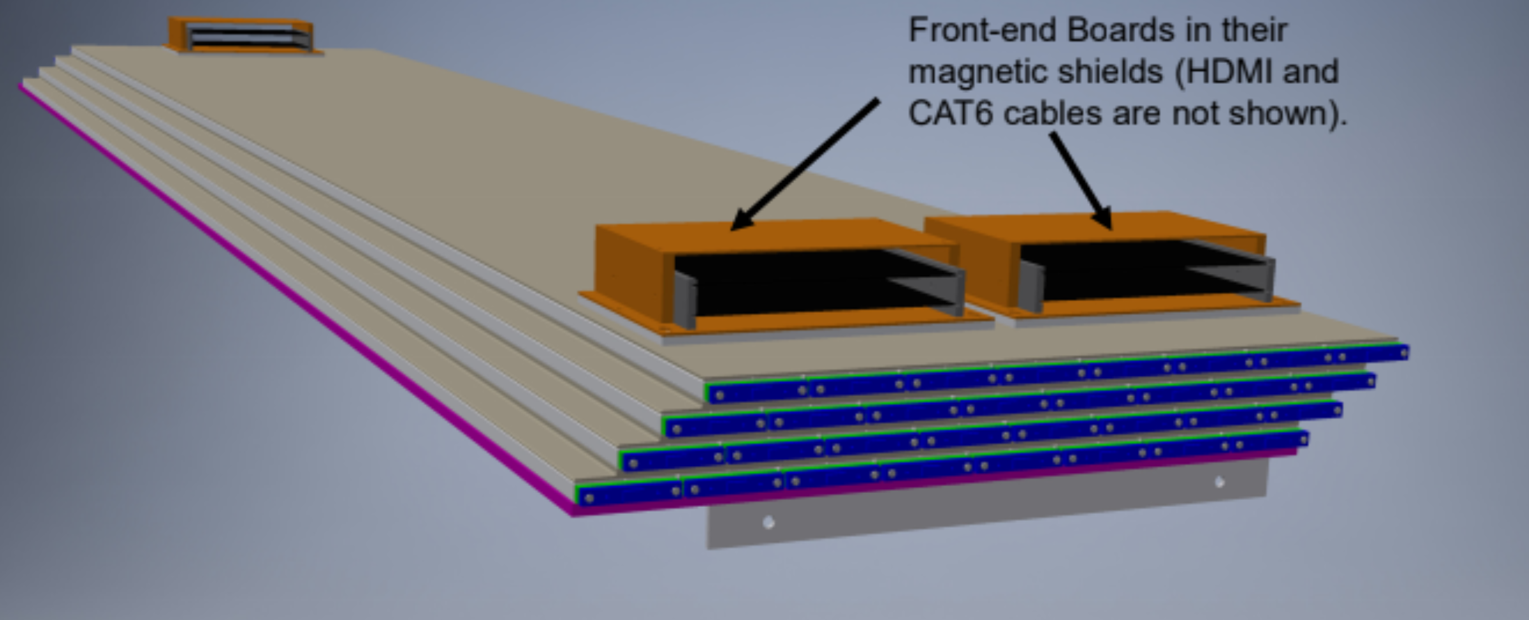}
    \caption{Schematic of a module, containing 32 dicounters in four layers.}
    \label{fig:my_label3}
\end{figure}

\subsection{Module Assembly and Quality Assurance}
Prior to module assembly, aluminum tape is wrapped around the
interface of the scintillator and FGB of each di-counter to insure
that di-counter is light-tight. A module is composed of four layers
of eight di-counters glued together with aluminum sheets in between
each layer, as shown in Fig.~\ref{fig:my_label3}. The aluminum is needed for mechanical integrity and to reduce electron punch-through. A vacuum bag is used to compress the module components while the epoxy sets. Once constructed, any detected light leaks in a module are sealed. The light yield and efficiencies of the di-counters in a module are measured with a cosmic ray test stand.

\subsection{Fabrication Status}
The completed CRV will consist of 5376 counters in 83 modules. Di-counter production is in progress at the University of Virginia with 58\% of the required di-counters built by September 2019. Module production began in early 2019, and six production modules have been assembled.


\begin{thebibliography}{99}



\bibitem{CLFV} 
  R.~H.~Bernstein and P.~S.~Cooper,
  ``Charged Lepton Flavor Violation: An Experimenter's Guide,''
  Phys.\ Rept.\  {\bf 532}, 27 (2013).

\bibitem{TDR} 
  L.~Bartoszek {\it et al.} [Mu2e Collaboration],
  ``Mu2e Technical Design Report,''
  arXiv:1501.05241.
  
\bibitem{SINDRUMII} 
  W.~H.~Bertl {\it et al.} [SINDRUM II Collaboration],
  Eur.\ Phys.\ J.\ C {\bf 47}, 337 (2006).

\bibitem{FiberPaper} 
  E.~C.~Dukes, P.~J.~Farris, R.~C.~Group, T.~Lam, D.~Shooltz and Y.~Oksuzian,
  JINST {\bf 13}, no. 12, P12028 (2018).


  
\bibitem{TestBeam} 
  A.~Artikov {\it et al.} [Mu2e Collaboration],
  Nucl.\ Instrum.\ Meth.\ A {\bf 890}, 84 (2018).
  

\end{thebibliography}
\end{document}